# AI-Driven Anonymization: Protecting Personal Data Privacy While Leveraging Machine Learning

Le Yang1*& Miao Tian1

Computer Information Science,Sam Houston State University,Huntsville，TX，USA,wesleyyang96@gmail.com

Master of Science in Computer Science,San Fransisco Bay University ,Fremont CA, USA,miao.hnlk@gmail.com

Duan Xin2

Accounting,Sun Yat-Sen University,HongKong,duanxin12314057@gmail.com

Qishuo Cheng3

Department of Economics,University of Chicago,Chicago, IL, USA,qishuoc@uchicago.edu

Jiajian Zheng4

Bachelor of Engineering,Guangdong University of Technology,ShenZhen,CN,im.jiajianzheng@gmail.com

Abstract: The development of artificial intelligence has significantly transformed people's lives. However, it has also posed a significant threat to privacy and security, with numerous instances of personal information being exposed online and reports of criminal attacks and theft. Consequently, the need to achieve intelligent protection of personal information through machine learning algorithms has become a paramount concern. Artificial intelligence leverages advanced algorithms and technologies to effectively encrypt and anonymize personal data, enabling valuable data analysis and utilization while safeguarding privacy. This paper focuses on personal data privacy protection and the promotion of anonymity as its core research objectives. It achieves personal data privacy protection and detection through the use of machine learning's differential privacy protection algorithm. The paper also addresses existing challenges in machine learning related to privacy and personal data protection, offers improvement suggestions, and analyzes factors impacting datasets to enable timely personal data privacy detection and protection.

CCS Concept:Choose Relevance:Security and privacy• Database and storage security•Database activity monitoring

---

[1] *These authors contributed equally to this work and should be considered co-first authors.*

## 1  INTRODUCTION

"Artificial intelligence technology is constantly being iterated and applied to more and more industries. Generative AI, which can create text and chat with users, presents a unique challenge because it can make people feel like they're interacting with a human. Anthropomorphism is the ascription of human attributes or personality to non-humans. People often anthropomorphize artificial intelligence (especially Generative AI) because it can create human-like outputs. Among them, information transmission activities based on artificial intelligence technology have received more and more attention. With the help of artificial intelligence technology to obtain information and transmit information, it can be more convenient and accelerate the realization of information interaction, industry marketing, user interaction, brand publicity, and advertising, and create more creative content. Artificial intelligence technology has brought great changes and more availability to everyone's daily life and receiving information channels. However, the collection of personal data is more and more extensive, which also makes the problem of personal data privacy and security more serious. Therefore, combined with the double-sided nature of artificial intelligence, this paper analyzes the advantages and disadvantages of intelligent data processing in personal data privacy, applies the machine learning differential privacy algorithm combined with intelligent data processing to the research, and realizes the risk prediction and protection of personal data. This serves as a reminder for everyone on how to use artificial intelligence to protect their information security more effectively."

## 2  MACHINE LEARNING AND PRIVACY PROTECTION

Machine learning is the core technology today and requires a lot of data when training models. How to protect this data cheaply and efficiently is an important issue. This chapter introduces machine learning and its privacy definitions and threats, summarizes the current situation in the field of privacy protection, analyzes advantages and disadvantages, and looks forward to possible research directions in the future.

### 2.1  Machine learning

Machine learning (ML) uses computers to effectively mimic human learning activities. It learns from existing data and produces useful models to make decisions about future behavior.

In traditional machine learning training, the data of all parties is first collected by the data collector, and then the data analyst conducts model training. This mode is called centralized learning. It can be seen that in the centralized learning mode, once the user has collected data, it is difficult to have control over the data, and it is unknown where and how the data will be used.

### 2.2  Privacy protection based on machine learning

The research on privacy protection in machine learning can be roughly divided into two main lines: the Federated Learning (FL) and homomorphic encryption (HE), and the perturbation method represented by differential privacy (DP). The encryption method not only encodes the plaintext into the ciphertext that only certain personnel can decode, but also ensures the confidentiality of the data in the process of storage and transmission. At the same time, the cipher text can be calculated directly and the correct result can be obtained by means of the security protocol. However, the data encryption process often involves a large number of calculations, which will produce huge performance overhead in complex cases, so it is difficult to implement in practical application scenarios. The main challenge for this method is to design a reasonable perturbation mechanism to better balance the privacy and availability of the algorithm.

### 2.2.1 Federated learning (FL)

Federated Learning (FL) is a machine learning setup. Many clients, such as mobile devices or entire organizations, work together to train a model. This process is coordinated by a central server, like a service provider. The key aspect is that the training data remains decentralized and distributed.

Federated learning allows training data to be shared between multiple participants without compromising their data privacy. Participants' data privacy can be protected by uploading only model parameters, such as gradients, instead of uploading training data. Federated learning enables a unified machine learning model to be trained from local data of multiple participants while protecting data privacy.

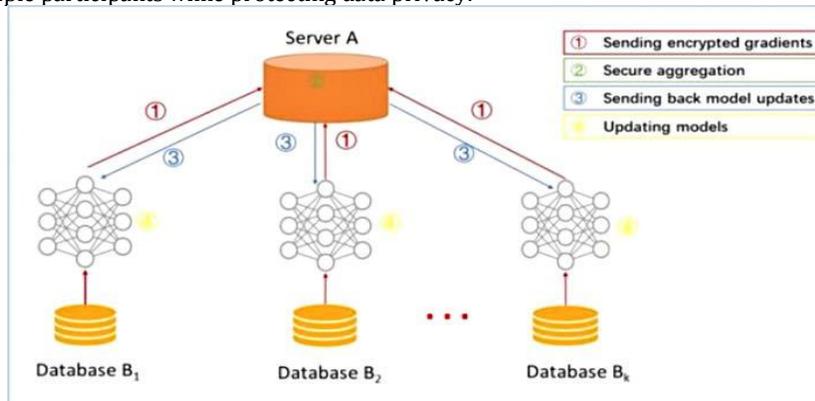

Figure 1: Federated learning algorithm model

Federated transfer learning is a data distribution situation in which horizontal federated learning and vertical federated learning do not match. There are also two methods of privacy protection in the federal learning mode: encryption and disturbance. However, federated learning is currently in the infancy of research and still faces many problems in both technology and deployment.

### 2.2.2 Homomorphic Encryption (HE)

Homomorphic encryption (HE) is a special encryption mode in cryptography. It enables us to send encrypted ciphertext to any third party for calculation without decrypting it beforehand. In other words, computations are performed on the ciphertext. Although the concept of Homomorphic Encryption first appeared in 1978, the first Fully Homomorphic Encryption framework that supports arbitrary operations on ciphertexts came later. Homomorphic encryption is an encryption method that supports computation on ciphertext.

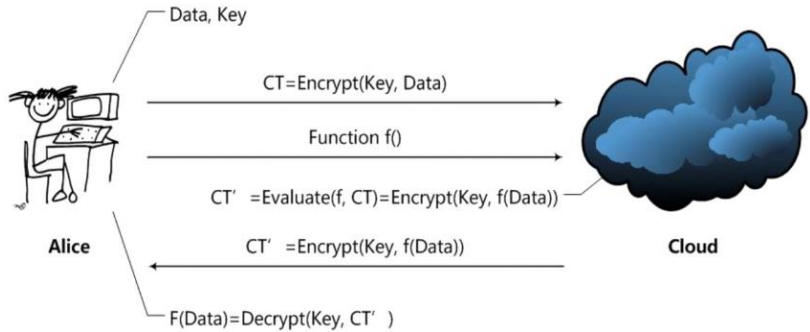

Figure 2: Schematic diagram of HE

The principle of homomorphic encryption technology is to provide a kind of encryption data processing function. That is, someone else can process the encrypted data, but the processing will not reveal any of the original content. At the same time, the user with the key decrypts the processed data and gets exactly the processed result. However, the encryption method calculation cost is not high, but the communication process designs a large number of security keys and other parameters, so the communication cost will be higher than the calculation cost.

### 2.2.3 Differential privacy protection algorithm (DP)

Differential privacy (DP) protection is achieved by adding noise to sensitive data. In federated learning, random noise is often added to the parameters that participants upload to the server in order to avoid backward inference of participants' private information.Compared with encryption methods, differential privacy mechanism is easier to deploy and apply in actual scenarios. Therefore, this paper focuses on the design of machine learning algorithm under differential privacy protection according to different data processing and analysis capabilities. for personal data privacy protection and risk prediction, differential privacy protection algorithm is the most commonly used model learning strategy for empirical risk minimization.

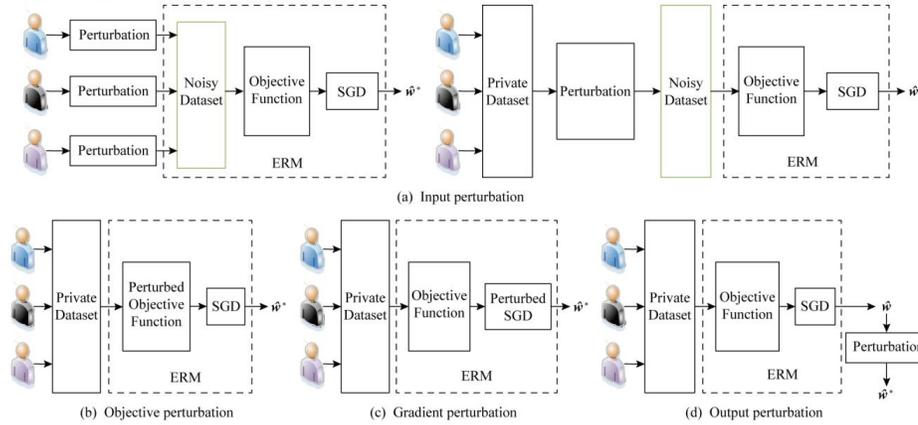

Figure 3: Empirical risk minimization for differential privacy protection

The empirical risk minimization of differential privacy protection algorithm is commonly used to solve the optimal model parameters by gradient descent (GD) based on iterative computation. Because of the simple structure of the traditional machine learning model, the objective function $J(w; D)$ make it as convex as possible in order to obtain a definite optimal solution. Due to the introduction of a large number of nonlinear factors, the objective function of deep learning model is often nonconvex, so it is easy to fall into the local optimal solution when solving. In addition, deep learning also has some problems, such as large number of parameters, many iterations, and slow algorithm convergence. Therefore, the privacy protection methods of the model in Figure 3 are quite different. Such algorithms have certain advantages in the protection of personal privacy data under the premise of intelligent driven anonymization, and localized differential privacy can prevent the server and the adversary from directly obtaining the original user data.

Based on the advantages and algorithm models of the above three algorithms, this paper finally chooses to complete the experimental demonstration part of this article by discussing the application process and scenario of machine learning under differential privacy protection to personal privacy data protection.

## 3 METHODOLOGY

### 3.1 Create Data set

The creation of a data set is the most important step for a differential privacy protection algorithm, where the input is the original data set and the output is a synthetic data set with the same shape (i.e. the same number of columns and the same number of rows). And the values in the synthesized data set during data creation have the same properties as the corresponding values in the original data set.

A histogram is a synthetic representation suitable for answering certain queries, but its shape is different from the raw data, so it is not synthetic data. So as a first step in obtaining synthetic data, we'll design a synthetic representation for a column of the original dataset that can answer the range query. A range query calculates the number of rows in a dataset whose values are within a given range.

For example, "How many participants were between the ages of 21 and 33?" Is a range query.

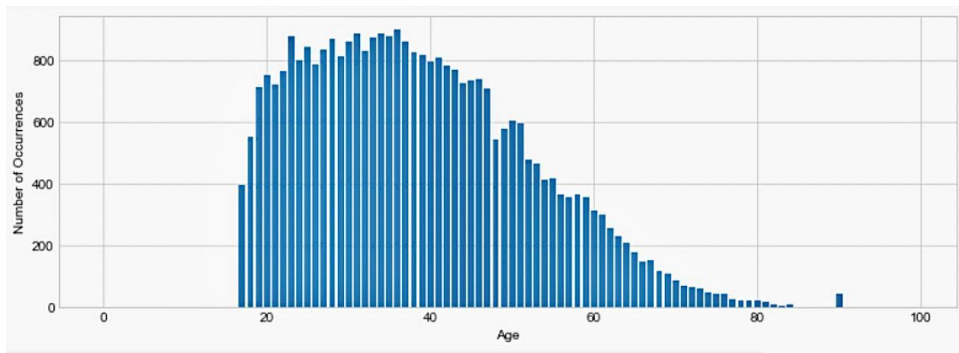

Figure 4: Data set query range results

Figure 4 The query defines a histogram bar for each age between 0 and 100 and uses a range query to count the number of people in each bar. The result looks very similar to the output of calling plt.hist data, since we are basically calculating the same result manually.

### 3.2 Add differential privacy

Suppose there is a stochastic algorithm M, S is the set of all possible outputs of M, and Pr[] represents the probability. For any two adjacent data sets. Two data sets The difference between two data sets is only 1 record if the probability distribution of two adjacent sets satisfies the following constraints:

$$P_r[M(D) \in S] \leq e^{\epsilon} P_r[M(d') \in S] \quad (1)$$

When the above conditions are met, it is said that algorithm M provides differential privacy protection, where $\epsilon$ is the differential privacy budget, which is used to ensure that one data record is added or less in the data set.

In this experiment, when adding differential privacy, I need to make the synthesized data compliant with differential privacy. Each count in the histogram is added separately, usually by Laplacian noise, by parallel combination, which satisfies $\epsilon$ \epsilon$\epsilon$- differential privacy.

Code:

```
epsilon = 1
dp_syn_rep = [laplace_mech(c, 1, epsilon) for c in counts]
```

Through post-processing, these results also satisfy ϵ \epsilonϵ- differential privacy.

### 3.3 Generate tabular data

Consider the composite representation as a probability distribution that estimates the underlying distribution from which the original data is obtained and from which a sample is drawn. Since we only consider one column and ignore all the others, this is called a marginal distribution (especially a one-way marginal distribution [1-way marginal]). So you need to count each histogram bar; We will normalize these counts so that they sum to 1 and then treat them as probabilities.

Finally, these probabilities are sampled from the distribution it represents by randomly selecting a bar of the histogram, weighted by probability. Our first step is to prepare the counts by making sure that no counts are negative and normalizing them to 1:

Code:

```
dp_syn_rep_nn = np.clip(dp_syn_rep, 0, None)
syn_normalized = dp_syn_rep_nn / np.sum(dp_syn_rep_nn)
np.sum(syn_normalized)
```

In the synthetic data above, if we plot the normalized count - which we can now think of as the probability of each corresponding histogram bar since they sum to 1 - we see a shape that looks very much like the original histogram (and, in turn, looks a lot like the shape of the original data). This is all to be expected - apart from their scale, these probabilities are just counting.

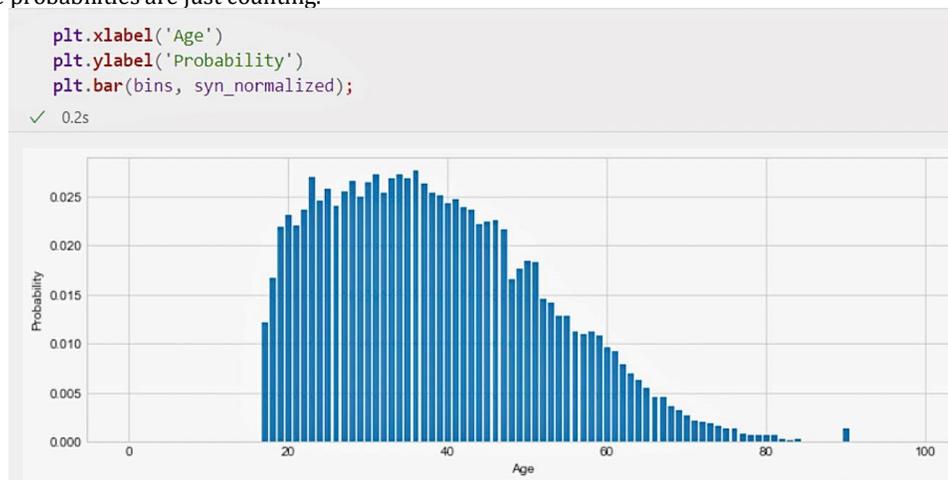

Figure 5: Data probability result histogram

The final step is to generate a new sample based on these probabilities. We can use np.random.choice, which allows you to pass in a list of probabilities associated with the choice given in the first argument (in the argument). It accurately implements the weighted random selection required for the sampling task. We can generate as many samples as we need without additional privacy costs because we have set the count to differential privacy.

### 3.4 Training model

This approach is simple, but since we consider each column in isolation, we will lose the correlation between the columns present in the original data. For example, in the data, age and occupation may be correlated (e.g., managers are more likely to be older than younger); If we consider each column in isolation, we would get the number of 18 year olds and managers right, but we could be very wrong about the number of 18 year old managers.

Another approach is to consider multiple columns together. For example, we can consider both age and occupation, and calculate how many 18-year-old managers there are, how many 19-year-old managers there are, and so on. The result of this modification process is a two-way marginal distribution. We will eventually consider all possible combinations of age and occupation - which is exactly what we did when we built the emergency table earlier! For example:

Code:

ct = pd.crosstab(adult['Age'], adult['Occupation'])

ct.head()

```python
ct = pd.crosstab(adult['Age'], adult['Occupation'])
ct.head()
✓ 0.1s                                                                                           Python
```

| Occupation | Adm-clerical | Armed-Forces | Craft-repair | Exec-managerial | Farming-fishing | Handlers-cleaners | Machine-op-inspct | Other-service | Priv-house-serv | Prof-specialty | Protective-serv | Sales | Tech-support | Transport-moving |
|---|---|---|---|---|---|---|---|---|---|---|---|---|---|---|
| Age | | | | | | | | | | | | | | |
| 17 | 23 | 0 | 14 | 1 | 9 | 40 | 2 | 129 | 8 | 10 | 3 | 87 | 1 | 3 |
| 18 | 55 | 0 | 17 | 6 | 14 | 50 | 17 | 152 | 4 | 10 | 5 | 115 | 2 | 8 |
| 19 | 102 | 0 | 40 | 12 | 24 | 65 | 30 | 166 | 3 | 18 | 3 | 112 | 8 | 16 |
| 20 | 117 | 0 | 35 | 15 | 23 | 81 | 41 | 139 | 3 | 28 | 9 | 108 | 14 | 24 |
| 21 | 121 | 0 | 59 | 18 | 25 | 51 | 51 | 142 | 4 | 30 | 7 | 93 | 16 | 14 |

Figure 6: Training model result

The disadvantage of considering both columns is that our accuracy will be lower. When we add more columns to the set under consideration (i.e., build an n nn to the edge, increasing the value of n nn), we see the same effect as for the contingent table - each count gets smaller, so the signal gets smaller relative to the noise, and our results are less accurate. We can see this effect by plotting the age histogram in the new composite data set; Note that it has roughly the right shape, but it's not as smooth as the raw data or the differential private count we used for the age column itself.

### 3.5 Experimental result

In this experiment, differential privacy protection algorithms play a crucial role in synthesizing a data set from the original data while ensuring privacy. The methodology starts with creating a synthetic data set that retains the shape and statistical properties of the original data.Specifically, for attributes like age, the process involves generating a histogram to represent the distribution of ages, ensuring that the synthetic data provides accurate answers to range queries without revealing individual data points.The introduction of noise, typically through mechanisms like Laplacian noise, into the synthetic representation ensures differential privacy by making it mathematically improbable to distinguish whether any individual's data is included in the data set.This process allows for the generation of synthetic data that can be used for various analytical purposes without compromising the privacy of the individuals represented in the original data set.Furthermore, the experiment highlights the challenges of maintaining correlations between different data attributes (e.g., age and occupation) when generating synthetic data.In conclusion, the experiment underscores the effectiveness of differential privacy protection algorithms in safeguarding individual privacy while providing a functional synthetic dataset for analysis, albeit with considerations regarding the accuracy and inter-attribute correlations in the synthetic data.

## 4 CONCLUSION

Differential privacy is a new definition of privacy protection proposed by Dwork et al in 2006 for the privacy leakage problem of statistical databases, and has now developed into the most advanced privacy protection method. Differential privacy protection can solve the two defects of traditional privacy protection model. Under this maximum background knowledge assumption, differential privacy protection does not take into account any possible background knowledge of the attacker, and it is built on a solid mathematical foundation, provides a strict definition of privacy protection and quantitative evaluation methods, so that the level of privacy protection provided by data sets under different parameter processing is comparable. These theories provide a way for decision makers to accurately control the level of privacy protection.

By adding differential privacy protection to the original data, this paper proves the application process of differential privacy algorithm based on machine learning in anonymizing personal data protection. Therefore, it can be seen that these algorithms can effectively reduce the risk of data leakage by applying mathematically strict privacy protection on the data set. It also allows researchers and institutions to use the data for important analysis and model training. As technology advances and privacy awareness increases, more innovative differential privacy technologies are expected to emerge, which not only provide more efficient data protection, but also ensure the usefulness of data and the accuracy of analysis. In addition, with the gradual improvement of regulations and policies, differential privacy protection algorithms are expected to become part of data processing standards, thus promoting the implementation and standardization of personal privacy protection worldwide.

**ACKNOWLEDGMENTS**

In the process of completing this research, I feel very honored to be able to draw on and quote the research results of Liu Bo and his collaborators. Their research has provided valuable support and guidance to my exploration in the fields of artificial intelligence, machine learning and personal privacy data protection. And I want to thank Dr. Liu Bo and his research team, In their paper "Integration and Performance Analysis of Artificial Intelligence and Computer Vision Based on Deep Learning. The core conclusions presented in Algorithms (arXiv preprint arXiv:2312.12872, 2023) provide a solid foundation for my research. This paper deeply discusses the integration and performance analysis of artificial intelligence and computer vision on the basis of deep learning algorithms, providing a new perspective for my understanding of artificial intelligence. In addition, Liu Bo's outstanding research has set an example in the field of machine learning and deep learning. His work has been instrumental in advancing the development of artificial intelligence, as well as providing a key context and framework for this article, so I would like to express my heartfelt gratitude to Liu Bo and his team.